\documentclass{sig-alternate-per}

\usepackage[margin=1in]{geometry}

\usepackage{amsmath}

\usepackage{framed}
\usepackage{url}

\begin{document}
\pagestyle{empty}

\title{Unbalanced Job Approximation using Taylor Series Expansion and Review of Performance Bounds}
\author{Alexander Thomasian \\
Thomasian \& Associates \\
Pleasantville, NY 10570 USA\\
alexthomasian@gmail.com
}
\date{}
\maketitle

\begin{abstract}
Unbalanced Job Approximation - UJA is a family of low-cost formulas 
to obtain the throughput of Queueing Networks - QNs with fixed rate servers
using Taylor series expansion of job loadings with respect to the mean loading.
UJA with one term yields the same throughput as optimistic Balanced Job Bound - BJB, 
which at some point exceeds the maximum asymptotic throughput. 
The accuracy of the estimated throughput increases with more terms in the Taylor series. 
UJA can be used in parametric studies by reducing the cost of solving large QNs by aggregating stations 
into a single Flow-Equivalent Service Center - FESCs defined by its throughput characteristic.
While UJA has been extended to two classes it may be applied to more classes by job class aggregation.
BJB has been extended to QNs with delay servers and multiple jobs classes by Eager and Sevcik,
throughput bounds by Eager and Sevcik, Kriz, 
Proportional Bound - PB and Prop. Approximation Bound - PAM by Hsieh and Lam and 
Geometric Bound - GB by Casale et al. are reviewed. 
\end{abstract}


\section{Introduction to Queueing Network Modeling}

{\it Queueing Networks - QNs} are used to model the realistic situation 
where customers (resp. jobs) require processing at more than one service station 
(resp. device at computer system). 
QNs have been used in many application domains such as manufacturing systems,
computer communication networks, and multiprogrammed computer systems.

While efficient solution methods have been developed for 
a subcategory of QN models known as product-form, 
the solution cost increases with the number of stations, 
number of job classes and the number of jobs.

QN models can be categorized to open, closed, and mixed.
In open QNs there are external job arrivals and completed jobs leave the system,
while in closed QNs completed jobs are immediately replaced with new jobs, 
as if there is an infinite backlog of jobs.
We consider product-form QNs, which lend themselves to an exact solution 
and can be analyzed relatively efficiently Kleinrock 1975 \cite{Klei75}.

The analysis of open product-form QNs with Poisson arrivals is trivial
in that each station can be analyzed separately Jackson 1957 \cite{Jack57}.
Closed PF QNs require the degree of concurrency and 
similarly to open QNs service demands or loadings, 
which is the mean total time jobs spend being served at service stations.

{\it Unbalanced Job Approximation - UJA} Thomasian and Nadji 1981 \cite{ThNa81} 
is a family of approximations based on Taylor series expansions
of job loading with respect to the average loading $X_0$. 
UJA starts with an upper bound on throughput, which equals 
the optimistic throughput of {\it Balanced Job Bounds - BJBs} Zahorjan et al. 1982 \cite{ZSEG82}.
UJA obtains more accurate throughput estimates by using a Taylor series expansion with respect to $X_0$. 
UJA was presented with BJB at SIGMETRICS'81 Zahorjan et al. 1981 \cite{ZSEG81},
but unlike BJB which was publicized in Lazowska et al. 1984 \cite{LZGS84},
UJA is little known and hence this publication.

UJA was applied in the analysis of the QN model of a computer system running CAD applications 
originated by users of {\it Time-Sharing Option - TSO} 
on an IBM mainframe with the MVS operating system. 
A performance study based on BEST/1 capacity planning tool \cite{Buz+78} 
is reported by Silvester and Thomasian 1981 \cite{SiTh81}.
Given large number of disks measurement results were used 
to aggregate disks with almost balanced utilizations,
while treating disks with very low utilizations as delay servers \cite{LZGS84}.


The paper is organized as follows.
Product-form QNs are described in Section \ref{sec:closed}.
The GF method for analyzing closed QNs is presented in Section \ref{sec:GF}.
Aggregation of balanced stations to a {\it Flow Equivalent Service Center - FESC}
is described in Section \ref{sec:balanced}.
This is followed UJA for a single and two job classes  
in Section \ref{sec:Taylor} and Section \ref{sec:twoclass}
based on Nadji and Thomasian 1981/1984 \cite{ThNa81,NaTh84}
Conclusions are drawn in Section \ref{sec:conclusion}.
Appendix I presents {\it Performance Bound Hierarchies - PBH} 
by Eager and Sevcik 1983,86 \cite{EaSe83,EaSe86}
Appendix II presents extensions to BJB by Kriz 1984 \cite{Kriz84}. 
Appendix III discusses asymptotic expansions with multiple bottlenecks by George et al. 2012 \cite{GeXS12}.
Appendix IV presents Geometric Bounds - GB by Casale et al. \cite{CaSM08}
and compares them with Proportional Bounds - PB by Hsieh and Lam \cite{HsLa87}.
Appendix V present Proportional Approximation Method - PAM 
for multichain QNs by Hsieh and Lam 19989 \cite{HsLa89}.
We have preserved the notation of the original papers 
in the Appendices to make it easer for readers to refer to them.

\vspace{5mm}
\section{Closed QNs with Product-Form Solutions}\label{sec:closed}

Product-form QNs allow stations with  single and multiple servers 
with exponential service times and FCFS scheduling Jackson 1957 \cite{Jack57}. 
Arrivals to open QNs are Poisson and jobs are routed from station 
to station according to probabilistic routing.
The analysis of such QN models is inexpensive, 
since each station can be analyzed separately as if it is subjected to Poisson arrivals.

The arrival rates of jobs to the $N$ stations is given by:
$$ \underline{\gamma} = ( \gamma_1, \gamma_2, \ldots , \gamma_N) $$
Job arrival rates to the $N$ stations, denoted by $\lambda_n, 1 \leq n \leq N$, 
are obtained by solving the set of linear equations:  

\begin{eqnarray}\label{eq:routing}
\underline{\lambda} = \underline{\gamma} + \underline{\lambda} P \mbox{ hence }
\underline{\lambda} = [I-P]^{-1} \underline{\gamma}.
\end{eqnarray}

Server utilization factor at the $n^{th}$ station is $\rho_n =\lambda_n \bar{x}_n / m_n$,
where $\bar{x}_n$ is the mean service time and $m_n$ the number of servers.
In the case $m_n =1$ and exponential service times $\bar{x}_n=1/\mu_n$,
the mean residence time at the station is that of an M/M/1 queue.
$$R_n = \bar{x}_n / (1- \rho_n) = (\mu_n - \lambda_n)^{-1}\mbox{ for }\rho_n < 1$$
The formula for M/M/m queues is given in \cite{Klei75}.   

\begin{framed}
\subsubsection*{Central Server Model - CSM}

CSM is a closed QN model of a multiprogrammed computer system proposed by Buzen 1973 \cite{Buze73},
which consists of a CPU and multiple disks.
Jobs alternate between CPU and disk processing until they are completed.
An efficient solution method known as the {\it Convolution Algorithm - CA} 
is presented in \cite{Buze73} to analyze CSM and closed product-form QNs in general.

The CPU is designated as the central station ${\cal S}_1$ 
and the $N-1$ disks as peripheral stations ${\cal S}_n, 2 \leq n \leq N$.
Given the state transition probabilities ${\cal S}_i \xrightarrow{p_{i,j}} {\cal S}_j$
the following transitions are applicable to CSM.
$$p_{1,n}, \hspace{2mm}2 \leq n \leq N, \hspace{3mm}p_{n,1}=1, \hspace{2mm} 2 \leq n \leq N$$

The self-transition $p_{1,1}=1 - \sum_{n=2}^N p_{1,n}$ implies the completion of a job
in a closed QN (or a job that leaves the system in an open QN).
The number of visits to the CPU is given by the geometric distribution \cite{Klei75}.
$$q_k = p_{1,1} (1-p_{1,1})^{k-1}, k \geq 1 \hspace{2mm}=\bar{k}=v_1 = 1/p_{1,1}.$$

The relative number of visits to the stations is obtained by solving 
$$\underbar{v}=\underbar{v}{\bf P}$$
It follows $v_n = p_{1,n} v_1 = p_{1,n}/p(1,1), 2 \leq n \leq N$.    

Given mean service time at ${\cal S}_n$ per visit is $\bar{x}_n$,
the mean loading per job is $X_n=v_n \bar{x}_n, 1 \leq n \leq N$. 
\end{framed}

Inputs required for the analysis of closed QNs 
are the number of jobs or the degree of concurrency,
probabilistic routings among stations which lead to the mean number of visits to the stations
and mean service times per visit, which yields loadings. 
Performance metrics of interest in modeling closed QNs are system throughput,
mean residence time, device utilizations and mean queue-lengths.  



Jackson theorem which only allowed service stations with exponential servers with FCFS scheduling 
was extended by the BCMP theorem to four types of stations Baskett et al. 1975 \cite{BCMP75}.

\begin{eqnarray}\label{eq:BCMP}
F_n(k) =
\begin{cases}
X_n^k     \hspace{10mm} \mbox{FCFS, PS, LCFSPR}                         \\
X_n^k/k   \hspace{9mm}  \mbox{Infinite-Server - IS}                     \\
X_n^k / \prod_{j=1}^k a(j)  \hspace{6mm} \mbox{Queue-dependent}
\end{cases}
\end{eqnarray}
In a queue-dependent station the service rate varies with the number of jobs:
$$F_n(k)=X_n^k / a(k),\mbox{  where  }a(k)= \mu(k) / \mu(1)$$
$a(k)$ is the ratio of the service rate with $k$ versus one job at the station.
Multiserver and IS or delay server queues are special cases with
$a(k)=k, 1 \leq k \leq m$ for $m$-server. For IS or delay servers $m=\infty$.
General service times are allowed for {\it Processor Sharing - PS} and IS service stations.

Buzen 1973 \cite{Buze73} developed the {\it Convolution - CA} algorithm 
to analyze closed QNs and applied it to CSM. 
We consider mostly fixed rate single server and delay servers in our discussion.

The solution of open product-form QN models is trivial.
Given the arrival rate to each station its mean residence time
can be obtained by solving the corresponding station independently.

Consider the processing of $K$ jobs in a closed QN with a single jobs with $N$ stations. 
Job service demands or loadings are the products of mean service time $\bar{x}_n$ 
and the mean number of visits ($v_n$) to station ${\cal S}_n$, i.e., $X_n = v_n \bar{x_n}$.

The steady-state state probability in a closed QN with $K$ jobs with 
$N$ fixed-rate single server-stations with job distribution:
$$\underline{k}=(k_1, k_2, \ldots , k_N) \mbox{ with }\sum_{n=1}^N = K.$$

\begin{eqnarray}\label{eq:GK}
P[\underline{k}]=
\frac{X_1^{k_1} X_2^{k_2} \ldots X_N^{k_N}} {G(K)}\hspace{2mm}
G(K)= \sum_{\underline{k} \in {\cal K}}  \prod_{n=1}^N  X_n^{k_n}.
\end{eqnarray}

The number of terms grows rapidly with the size of the network
such that a direct summation of multinomial expressions may be computationally intractable.
Multiple job classes or chains yield a more realistic representations of QN models, 
where each class has its own routings probabilities and 
service times per visit and hence different loadings at the stations.  
Given $R$ job classes and $k_r , 1 \leq r \leq R$ jobs in class $r$, the number of states is. 
$$\prod_{r=1}^R \binom{K_r + N-1}{N-1}$$ 

There are several efficient computational methods to analyze product-form closed QNs.
The {\it Convolution Algorithm - CA} was proposed in the context of CSM 
with a single job class Buzen 1973 \cite{Buze73} 
and was extended as part of the BEST/1 capacity planning package by Buzen et al. \cite{Buz+78}.

IBM's Reiser and Kobayashi 1975 \cite{ReKo75} used generating functions, see e.g., \cite{Klei75}, 
in extending BCA to multiple job classes.
The application of GFs to analyze closed QNs is presented in a tutorial manner in
Williams and Bhandiwad 1976 \cite{WiBh76}, Thomasian and Nadji 1981 \cite{ThNa81},
and Trivedi 2002 \cite{Triv02}, 

\section{Generating Functions for Analyzing Product-Form QNs with a Single Job Class}\label{sec:GF}

The GF for $n^{th}$ QN station for $1\leq n \leq N$ is given by:

\begin{eqnarray}\label{eq:GF}
x_n(t) = \sum_{k=0}^\infty F_n (k) t^k,
\end{eqnarray}
where $F_n (k)$ was given by Eq. (\ref{eq:BCMP}).

For stations with fixed-rate single servers $F_n(k)=X_n^k$.

\begin{align}\label{eq:SS}
x_n (t) &= \sum_{k \geq 0} (X_n t)^k = 1 + (X_n t) + (X_n t)^2 + \ldots \\ 
\nonumber
&= (1-X_n t)^{-1}, |X_n t| < 1 .
\end{align}
 
If the $n^{th}$ station is IS: 

\begin{eqnarray}
x_n (t) = 1 + X_n t + \frac{(X_n t)^2}{2!} + \dots = e^{X_n t}. 
\end{eqnarray}

The GF for the QN is the product of the GFs for individual stations.

\begin{eqnarray}\label{eq:gt}
g(t) = \prod_{n=1}^N x_n(t).
\end{eqnarray}

which can be rewritten as:

\begin{eqnarray}
g(t) = 1 + G(1)t + G(2)t^2 + \ldots + G(K)t^K + \ldots,
\end{eqnarray} 

In the case of single-server stations with $N=2$ stations as $K$ increases:        
$$
\begin{cases}
G(1)= X_1+ X_2,                                   \\
G(2)=X_1^2 + X_2^2 +X_1 X_2,                      \\
G(3)=X_1^3 + X_1^2 X_2 + X_1 X_2^2 + X_2^3 ,      \\
\vdots 
\end{cases}
$$ 

Given  
$$g_n(t) = g_{n-1} (t)x_n(t), 1 \leq n \leq N$$
and noting that $x_n (t)  = (1- X_n t)^{-1}$ it follows

$$ g_n (t) = g_{n-1} (t) + X_n g_n (t), \hspace{5mm} 1 \leq k \leq K, 1 \leq n \leq N $$
Equating the coefficients of $t^k$ on both sides we have: 

\begin{eqnarray}
\boxed{G_n (k) = G_{n-1} (k) + X_n G_n (k-1),}  \\
\nonumber
1 \leq k \leq K, 1 \leq n \leq N.
\end{eqnarray}

The utilization of a single server at station $n$ is 
the sum of state probabilities with at least one job at the station:

\begin{eqnarray}
\boxed{U_n (K) = X_n T(K),}  
\end{eqnarray}

The system throughput follows from Little's formula \cite{Klei75}:

\begin{eqnarray}\label{eq:TK}
\boxed{T(K) = \frac{U_n (k)}{X_k} = \frac{G(K-1)}{G(K)}.}
\end{eqnarray}
If $X$s are expressed in seconds $T(K)$ with be jobs/second.

The mean queue length at the $n^{th}$ single-server station is:

\begin{eqnarray}
\boxed{Q_n (K) = \frac{1}{G(K)} \sum_{k=1}^K X_n^k G(K-k).}
\end{eqnarray}

CA to compute the normalization constant $G(K)$ in a single class QN is:

\begin{framed}
Initializations:
$$
\begin{cases}
G_0(k) = 0 \hspace{5mm} 1 \leq k \leq K.  \\
G_n(0) = 1 \hspace{5mm} 0 \leq n \leq N
\end{cases}
$$
for $k=1$ to $K$ do    \newline
for $n=1$ to $N$ do    \newline
$G_n (k) = G_{n-1} (k) + X_n G_n (k-1)$ \newline
end                    \newline
end                  
\end{framed}

\subsection{Aggregation of Balanced Stations}\label{sec:balanced}

Given that a subset $M \leq N$ of the QN stations have balanced loads, i.e., $X_n=X, 1 \leq n \leq M$,
it follows from Eq. (\ref{eq:gt}) that:

\begin{eqnarray}\label{eq:balanced}
v(t) = [ 1 + X t + (X t)^2 + \dots ]^M = ( 1 - X t)^{-M}
\end{eqnarray}  

Applying the Binomial Theorem we have:

\begin{eqnarray}
v(t) = \sum_{k=0}^\infty \binom{M+k-1}{k} (Xt)^k.
\end{eqnarray}

The GF for the aggregate station can be written as:

\begin{eqnarray}
v(t) = \sum_{k=0}^\infty (M X t)^k / \prod_{j=1}^k a(j), \hspace{2mm} a(j) =\frac{ j M }{M+j-1}.
\end{eqnarray}

The throughput of the using Eq. (\ref{eq:TK}) is:

\begin{eqnarray}\label{eq:Tkbal}
T(k) = \frac{ \binom{M+k-2}{k-1} X^{k-1) } }                                 
{\binom{M+k-1}{k}X^k}    
\end{eqnarray}

It follows that job throughout with $j$ jobs is:
\begin{eqnarray}     
\boxed{ T(k) = \frac{k}{M+k-1}\frac{1}{X} }
\end{eqnarray}

As $k \rightarrow \infty$ the maximum throughput is $T_{max} = 1/X$

We can use the symmetry due to balancedness to obtain the mean residence time 
at single-server stations using the key equation in {\it Mean Value Analysis - MVA} 
Reiser and Lavenberg 1980 \cite{ReLa80} which is based on the Arrival Theorem 
Lavenberg and Reiser 1980 \cite{LaRe80}. 

In a QN with $k$ jobs a job arriving at a station sees a mean queue-length with one job less. 
The $k-1$ jobs are equally distributed among $M$ stations, hence:

$$r_n(k) = \bar{x}_k \left( 1+ \frac{k-1}{M} \right) $$
Multiplying both sides by $v_k$: 
\begin{eqnarray}
\boxed { R_n(k) = X_n \left( 1 + \frac{k-1}{M} \right). }
\end{eqnarray}

The mean residence time in the system is obtained by multiplying by $M$.
$$R(k)=M r(k)= (M+k-1)X,$$
from which the throughput $T(k)=k/R(K)$ as given by Eq. (\ref{eq:Tkbal}) follows. 

The aggregation of IS or delay station in a QN is particularly simple,
so that all such stations whose indices are in the set ${\cal I}$ 
can be replaced by a single IS station with service demand 
$$\boxed{ X_I = \sum_{i \in {\cal I}} X_i .}$$

\section{Aggregation of Unbalanced Stations in a Single Class}\label{sec:Taylor}

We use the throughput of an FESC with balanced stations as a starting point 
to obtain the throughout of an FESC with unbalanced stations.
Defining the mean service demand over $M$ fixed-rate single-servers stations: 
$$X_0 = \sum_{n=1}^M X_n / M.$$
The deviation of service demands from the mean and its moments are as follows:

\begin{eqnarray}
e_n = \frac{X_n - X_0}{X_0}, \hspace{2mm} 1 \leq n \leq M \mbox{   and   }E_j = \sum_{n=1}^M e_n^j.
\end{eqnarray}   

The GF of the $n^{th}$ station can be given as its Taylor series expansion as follows:

\begin{eqnarray}
x_n (t) = x_0 (t) + (X_n - X_0)x_0^{(1)} (t) +  \\
\nonumber
\frac{1}{2!} (X_n-X_0)^2 x_0^{(2)} (t) + \ldots ,
\end{eqnarray}

Making the substitution $(X_n - X_0) = e_n X_0$ 
and noting that that the $j^{th}$ derivative with respect to $t$:
$$x_0 ^{(j)} (t) = j! t^j [x_0 (t)]^{j+1}$$

\begin{eqnarray}
x_n (t) = x_0 (t) + (e_n X_0 t) x_0^{(2)}(t) +  \\
\nonumber
(e_n X_0 t)^2 x_0^{(3)}(t)  + (e_n X_0 t)^3 x_0^{(4)} (t) + \ldots 
\end{eqnarray}

Substituting $x_n(t)$ into Eq. (\ref{eq:balanced}) and noting that $E_1=0$ we have

\begin{eqnarray}\label{eq:vt}
v(t) = x_0^M (t) + \frac{1}{2} (X_0 t)^{M+2} E_2 x_0^{M+2} (t) + \\
\nonumber
\frac{1}{3} (X_0 t)^{M+3} E_3 x_0^{M+3} (t) + \ldots  
\end{eqnarray}

In the above expression $x_0^(M) (t)$ represents the GF for $M$ balanced stations.
It follows from the definition of $E_j$'s that this series will converge  
when the stations are not highly unbalanced and $e_n$s are small. 

The coefficient of $t^k$ in Eq. (\ref{eq:vt}) is given as:

\begin{framed}
\begin{eqnarray}\label{eq:vk}
V(k) = V_0 (k) + \frac{X_0^2}{2} V_0(k-2) E_2 \\
\nonumber
+ \frac{X_0^3}{3} V_0 (k-3) E_3 + \ldots 
\end{eqnarray}
\end{framed}
$V_0 (k =\binom{M+k-1}{k} X_0^k $ is the coefficient of $t^k$ in $X_0^M (t)$.

We can rewrite Eq. (\ref{eq:vk}) as 

\begin{eqnarray}
V(k) = V_0 (k) [1+ R_0 (k)], \mbox{   where   } \\
\nonumber
R_0 (k) \sum_{j=2}^\infty \frac{E_j}{j}
\left[ \prod_{i=0}^{j-1} \frac{k-i}{M+i} \right]
\end{eqnarray}

The throughput of the subnetwork with $k$ jobs is given as 

\begin{eqnarray}
T(k)=\frac{V(k-1)}{V(k)} = T_0(K) \frac{1+R_0 (k-1)}{1 + R_0 (k)} 
\end{eqnarray}
where $T_0 (k)$ is the throughput of the $M$ balanced stations.

We can obtain a family of approximations for $T(k)$ according 
to the number of terms in the summation for $R_0 (k)$.
Maintaining multiple terms in Eq. (\ref{eq:vk}):

\begin{eqnarray}\label{eq:2term}
V(k) = V_2 (k) + V_0(k) R_2 (k)
\end{eqnarray}

\begin{small}
\begin{eqnarray}\label{eq:V2k}
\nonumber
V_2 (k) = V_0(k) 
\left[
1 +\frac{1}{2}
\frac
{k(k-1)}{M(M-1)} 
E_2 +
\frac{1}{3} 
\frac
{k(k-1)(k-2)}
{M(M+1)(M+2)} E_3 
\right]
\end{eqnarray}
\end{small}

\begin{eqnarray}\label{eq:R2k}
\nonumber
R_2 (k) = \sum_{j=4}^\infty \frac{E_j}{j}  
\prod_{i=0}^{j-1} \frac{k-i}{M+i} 
\end{eqnarray}

The expression for throughput with with one and two summation terms are:

\begin{eqnarray}\label{eq:T1k}
T_1 (k)=
T_0 (k) 
\left[
\frac{  1 - \frac{k-1}{M+1} E_2 }
{1+ \frac{k(k-1)}{2 (M+1)} E_2  }
\right]
\end{eqnarray}

\begin{eqnarray}\label{eq:T2k}
T_2 (k) = T_0 (k) 
\left[ 
1- 
\frac{ 
\frac{k-1}{M(M+1)} 
\left( E_2 +\frac{k-2}{M+2}  E_3 \right) 
}
{1+ 
\frac{k(k-1)}{M(M+1)} 
\left( \frac{1}{2} E_2 +\frac{1}{3} \frac{k-2}{M+2} E_3 \right)   
}
\right]
\end{eqnarray}


Given the coefficient of variation $c$, 
which is the standard deviation divided by the mean and the coefficient of skewness, 
$\beta$ which is the third central moment divided by $c^3$
we have $E_2 = Mc^2 $ and $E_3 = M \beta c^3$.                              
\footnote{\url{https://grapherhelp.goldensoftware.com/WTOPICS/WKS_Skew.htm}}            
Eq. (\ref{eq:T2k}) and Eq. (\ref{eq:T1k}) can be rewritten using this notation as: 

\begin{eqnarray}
\boxed{ T_1 (k) = T_0 (k) 
\left[ 1-
\frac{
\frac{k-1}{M+1} c^2 }
{ 1 + \frac{k(k-1)}{2(M+1)} c^2  }
\right].
}
\end{eqnarray}

\begin{eqnarray}
\boxed{
T_2 (k) = T_0 (k) 
\left[ 1 - \frac{ \frac{k-1}{M+1} c^2 \left( 1 + \frac{k-2}{M+2} c \beta \right) }
{ 1 + \frac{k(k-1)}{M+1} c^2 \left( \frac{1}{2} + \frac{k-2}{3(M+2)} c \beta \right) }
\right]
}
\end{eqnarray}

In \cite{ThNa81b} we determine the relative errors: $T_0(k)/T_2(k) -1$ (resp. $T_1(k)/T_2(k) -1$) 
for varying $c$ (resp. $c$ and $\beta$), with respect to $T_2(k)$ since it is quite accurate.

$T_0 (k)$ which is the throughput based on the average service demand $X_0$ 
yields the maximum throughput as shown in \cite{ZSEG82}. 

Given that there are $m_n$ servers at station $n$ with mean service demands $X_n$
it follows from the fact the utilization factor should not exceed one:
$$U_n (k) = \frac{T(k) X_n}{m_n} < 1$$ 

It is shown in \cite{MuWo74} that the maximum throughput of a closed QN 
is given as the minimum of service station throughputs, 
hence as $k$ is increased $T(k)$ is bounded as:

\begin{eqnarray}\label{eq:ABAp}
\boxed{
T_{max} = \mbox{min} \left( \frac{m_1}{X_1}, \frac{m_2}{X_2}, \ldots, \frac{m_N}{X_N} \right)
\geq T(k), k \geq 1. }
\end{eqnarray} 

Since we are concerned with single-server queues $m_n=1,\forall{n}$ and $T_{max}=1/D_{max}$.

In fact the throughput for lower values of $k$ the throughput is bounded 
by a line that connects the origin and $T(1) = [\sum_{n=1}^N X_n]^{-1}$.
Throughput convexity is shown by Dowdy et al. 1984 \cite{DEGS84}
for QN's with fixed rate and delay servers. 

Optimistic (resp. pessimistic) {\it Balanced Job Bound - BJB} analysis of QNs 
utilize the average (resp.  maximum) service demand
yielding an upper (resp. lower) bounds to throughput \cite{ZSEG82},\cite{LZGS84}
Since the optimistic BJB bound may exceed $T_{max}$
it should only be used up to the point it intersects $T_{max}$.

\begin{framed}
\begin{eqnarray}\nonumber
\frac{K}{(K+N-1)D_{max}} \leq T(k) \leq                   \\ 
\nonumber
\mbox{min} \left( \frac{1}{D_{max}},\frac{K}{(K+N-1)D_{avg}} \right).
\end{eqnarray}
\end{framed}

It follows easily from Eq. (\ref{eq:ABAp}) that when the sum of service demands is fixed
then the throughput is maximized when the service demands at the stations equal the mean.

From a pragmatic viewpoint there is a drop from the maximum throughput
as the {\it MultiProgramming Level - MPL} is increased in virtual memory systems.
This is due to increased paging which may ensue in a severe performance degradation.
The frequency of paging is given by the lifetime curve,
which is the time between page faults and the number of page frames dedicated to a program
(see e.g., Figure 9.5 in \cite{LZGS84}).
As MPL is increased less page frames can be allocated to each program 
A QN model is used to obtain the throughput characteristic vs the MPL,
which show a drop beyond a certain maximum throughput.
Increasing the MPL further may result in a severe degradation 
in throughput referred to as thrashing Denning 1970 \cite{Denn70}.

\section{Aggregation of Balanced Stations with Two Classes}\label{sec:twoclass}

The aggregation of balanced stations for a single class is extended 
to two job classes in the Appendix of \cite{ThNa81b} and Nadji and Thomasian 1984 in \cite{NaTh84}.
There are $N$ stations and $K$ and $L$ jobs in two classes 
with service demands given by $X_n=X, 1 \leq n \leq N$ and $Y_n=Y, 1 \leq n \leq N$. 
The equilibrium state probability distribution is given as:

\begin{eqnarray}
P[{\bf k},\underline{\ell}]= \frac{1}{G(K,L)} \prod_{n=1}^N F_n (k_n,\ell_n).
\end{eqnarray}

For single-server stations with FCFS, PS, and LCFSPR queueing disciplines:
In the first case both job classes should 
have the same exponentially distributed service times.

\begin{eqnarray}
F_n (k_n,\ell_n) = \frac{ (k+\ell)! } { k_n! \ell!} X_n^{k_n} Y_n^{\ell_n}.
\end{eqnarray}

The GF for each station is given by:

\begin{align}\label{eq:htu}
h_n (t,u) 
&= \sum_{k=0}^\infty \sum_{\ell=0}^\infty 
\frac{(k+\ell)!}{k!\ell!} (X_n t)^k (Y_n u)^\ell     \\ 
\nonumber
&= (1- X_n t + Y_n u)^{-1} \mbox{  provided  } |X_n t + Y_n u | < 1.
\end{align}

In the case of IS stations 
$$F_n (k, \ell) = \frac{ (X_n t)^k}{k!} \frac { (Y_n u)^\ell }{\ell!}$$ 

\begin{align}
h_n (t,u) = 
\sum_{k=0}^\infty 
\sum_{\ell=0}^\infty 
\frac{ (X_n t)^k}{k!}  \frac { (Y_n u)^\ell }{\ell!} = \\
\nonumber
\sum_{m=0}^\infty (X_n t  + Y_n u)^m = e^{X_n t + Y_n u}.
\end{align}

The GF for the $M \leq N $ stations to be aggregated is

\begin{eqnarray}
\nonumber
v(t,u) = \prod_{i=1}^M h_i (t,u).
\end{eqnarray}

When the QN is balanced
$X_n = X, Y_n = Y, 1 \leq n  \leq N$.
The GF for the $M$ stations to be aggregated is:

\begin{eqnarray}
v(t,u) = [x(t,u)]^M = (1- X t - Y u)^{-M}              \\
\nonumber
\sum_{j=0}^\infty \binom{M+j-1}{j}(X_0 t + Y_0 u)^j. 
\end{eqnarray}

\begin{eqnarray}\label{eq:A13}
v(t,u)= \sum_{k=0}^\infty \sum_{\ell=0}^\infty V(k,\ell) (M X t)^k (M Y u)^\ell.
\end{eqnarray}

Equating the coefficients of $t^k u^\ell$ of $v(t,u)$

\begin{eqnarray}\label{eq:A15}
V(k,\ell)= \binom{M+k+\ell-1}{k+\ell}\binom{k+\ell}{k}.
\end{eqnarray}

The throughputs in the two classes are given as:

\begin{eqnarray}
T^{(1)} (K,L) = \frac{ V(K-1,L) }{ V(K,L) } = \frac{K}{M+K+L-1}.\frac{1}{X_0}    \\
\nonumber
T^{(2)} (K,L) = \frac{ V(K,L-1) }{ V(K,L) } = \frac{L}{M+K+L-1}.\frac{1}{Y_0}
\end{eqnarray}

If the QN is not balanced we expand the GF for ${\cal S}_n$ around the mean loadings $X_0$ and $Y_0$:

\begin{eqnarray}
h_n (t,u) =\sum_{j=0}^\infty \frac{1}{j!} \hspace{45mm}         \\
\nonumber
\left[ \left(                                                  
(X_n - X_0)\frac{\partial}{\partial X_n} +
(Y_n - Y_0)\frac{\partial}{\partial Y_n} \right)^j h_n(t,u) \right] =  \\
\nonumber
\sum_{j=0}^\infty \left[ (X_n - X-0)t +(Y_n - Y_0)u \right]^j h_0^{j+1} (t,u).
\end{eqnarray}

\begin{eqnarray}
v(t,u) / h_0^M (t,u) = \hspace{45mm} \\ 
\nonumber
1 + \frac{1}{2}  h_0^2 (t,u) \sum_{n=1}^M [ (X_n-X_0)t +  (Y_n-Y_0) u]^2 + \\ 
\nonumber
\frac{1}{3} h_0^3 (t,u) \sum_{n=1}^M [ (X_n-X_0)t +  (Y_n-Y_0) u]^3 + \ldots 
\end{eqnarray}

Equating the coefficients of $t^k u^\ell$

\begin{small}
\begin{eqnarray}
V(K,L) = V_0 (K,L) +
\end{eqnarray}
\vspace{-4mm}
\begin{eqnarray}
\nonumber                   
X_0^K Y_0^L \sum_{i=2}^\infty  \frac{1}{i} \sum_{j=0}^i 
\left[ \binom{M+K+L-1}{K+L-1} \binom{K+L-i}{K-j}\binom{i}{j} E_{j,i-j} \right]
\end{eqnarray}
\end{small}

$$E_{i,j} = \sum_{n=1}^M 
\left( \frac{X_n - X_0}{X_0} \right)^i
\left( \frac{Y_n - Y_0}{Y_0} \right)^j$$

The first order approximation can be expressed as:

\begin{eqnarray} 
V_1 (K,L) = V_0 (K,L) [ 1 + \frac{K(K-1)}{2M(M+1)} E_{2,0} + \\
\nonumber
\frac{L(L-1)}{2M(M+1)} E_{0,2} 
\nonumber
+\frac{KL}{M(M+1)}E_{1,1} ].
\end{eqnarray}

Given that $c_X$, $c_Y$, and $c_{X,Y}$ are the coefficients of variation and correlation for $X_n$ and $Y_n$ then:

\begin{eqnarray}
\nonumber
E_{2,0} = M c^2_X \hspace{2mm} E_{0,2} = M c^2_Y \hspace{2mm} E_{1,1} = M c_{X,Y} c_X c_Y.
\end{eqnarray}

\begin{eqnarray}
T_1^{(1)} (K,L) = T_0^{(1)} (K,L)  
\end{eqnarray}

\begin{tiny}
\begin{eqnarray}
\nonumber
\frac
{ 1- \frac{1}{2(M+1)} [ K(K-1)(K-2) c^2_X + L(L-1)c^2_Y +2(K-1) L c_X c_Y c_{X,Y} ] }
{1 + \frac{1}{2(M+1)} [    K(K-1)c^2_X + L(L-1) c^2_Y  + 2 K L c_X c_Y c_{X,Y} ]  }
\end{eqnarray}
\end{tiny}

\subsection*{Accuracy of Unbalanced Job Approximation}

A small experiment to assess the accuracy of UJA is given below.
The service demands at the $N=5$ stations are as follows:
$$\underline{X}= \{0.25, 0.23, 0.19.17, 0.15\},$$
such that $X_0 = 0.2$. $c=0.179$, and $\beta=0.079$.  
Exact results were obtained using CA.

\begin{table}[h]
\begin{tabular}{|c|c|c|c|c|}\hline
$k$      &$T_0(k)$    &$T_1(k)$    &$T_2(k)$  &$T(k)$  \\ \hline   
1        &1.0000      &1.0000      &1.0000    &1.0000  \\
2        &1.6667      &1.6578      &1.6578    &1.6578  \\
3        &2.1429      &2,1204      &2.1203    &2.1203  \\
4        &2.5000      &2.4612      &2.4611    &2.4610  \\
5        &2,7778      &2,7215      &2.7212    &2.7209  \\
6        &3.0000      &2.9259      &2.9254    &2.9245  \\ \hline
\end{tabular}
\caption{\label{tab:1}Comparison of approximate and exact throughputs for a single job class.}
\end{table}

A larger experiment with 100,000 random service demands in a QN with $N=18$ stations is reported in \cite{NaTh84}.
More than 99\% of the results obtained by $T_1$ and $T_2$ throughputs 
are within 10\% accuracy for the range of number of jobs considered. 
$T_0$ is accurate within 15\% in 97\% of the examples when average utilization was 40\%.
An experiment similar to Table \ref{tab:1} with two job classes is reported in \cite{NaTh84}.

\section{Conclusions}\label{sec:conclusion}

UJA - Unbalanced Job Approximation - UJA is a family of low-cost methods 
starting with an upper bound for the throughput based on the mean service demand $X_0$,
but more accurate throughput estimates are obtained using additional terms in a Taylor series expansion,     

Given a large number of stations, instead of aggregating all stations at once, 
aggregation can be applied to groups of single-server stations with close utilizations.

Given that UJA is restricted to two classes
and that it is difficult to extend to more classes,
more job classes can be dealt with by clustering multiple chains into one,
see e.g., deSouza et al. 1986 \cite{deLM86}. 
The error introduced by applying such methods is discussed in Cheng and Muntz 1986 \cite{ChMu96}.

Performance bound hierarchies for single and multiple job classes 
are given by Eager and Sevcik \cite{EaSe83,EaSe86}.
Kerola 1896 \cite{Kero86} is a less influential work on this topic. 

\section*{Appendix I: Performance Bound Hierarchies - PBH}

Eager and Sevcik 1983 \cite{EaSe83} extended BJB to {\it Performance Bound Hierarchies - PBH}
Consider a closed QN with $K$ stations, $N$ job and loadings $L_k, 1 \leq j \leq K$
According to the arrival theorem \cite{LaRe80}
\begin{eqnarray}\label{eq:e1}
R_k(N) = L_k [1 + \bar{n}_k (N-1)]
\end{eqnarray}
where $\bar{n}_k (N-1) $ is the mean number of jobs at station $k$.
It follows the mean residence time of jobs is $R(N)=\sum_{k=1}^K R_k (N).$ 
An application of Little's result yields.

\begin{eqnarray}\label{eq:e4}
\bar{n}_k (N-1) = \frac{R_k (N_1) }{Z+ R(N-1) } (N-1)
\end{eqnarray}

Substituting $n_k(N-1)$ in Eq. (\ref{eq:e1}) 

\begin{eqnarray}\label{eq:e5}
R_k^{(i)} (N) = L_k 
[ 1 + \frac{R_k^{(i)} (N_1) }
{Z+ R^{(i-1) }(N-1) }]  (N-1)
\end{eqnarray}

The optimistic hierarchy starts with the initialization:

\begin{eqnarray}\label{eq:E7}
R_{k\mbox{ opt}}^{(0)} = \frac{1}{K} \mbox{max}[N L_B - Z,1 ]
\end{eqnarray}
where $b$ is the index of the bottleneck station. It follows:

\begin{eqnarray}\label{eq:E8}
R_{k\mbox{ opt}}^{(0)} = \frac{1}{K} \mbox{max}[N L_b -Z,1 ]
\end{eqnarray}

Using $a(N) = \mbox{max}[N L_b -Z,1] $

\begin{eqnarray}\label{eq:E10}
R_{\mbox{opt}}^{(1)} = 1 +\frac{1}{K} \left( \frac{a(N-1)}{Z+a(N-1)}\right) (N-1). 
\end{eqnarray}

When $Z=0$ and $S=\sum_{k=1}^K L^2_k $

\begin{eqnarray}\label{eq:E12}
R^{(2)}_{\mbox{opt}} (N) = 1 +S(N-1).
\end{eqnarray}

For the pessimistic hierarchy we have

\begin{eqnarray}
R^{(0)}_{k \mbox {pess}} =
\begin{cases}
N \mbox{ for }k = b              \\
0 \mbox{ for }k \neq b 
\end{cases}
\end{eqnarray}

\begin{eqnarray}\label{eq:E14}
R^{ (0) }_{\mbox{pess}} (N) = N,
\end{eqnarray}

For $k \neq b$ $R_k{\mbox pess }^{(1)} = L_a$ and for $k=b$

\begin{eqnarray}\label{eq:E16}
R_k{\mbox pess }^{(1)} (N) = 1 + L_b \left( \frac{N-1}{Z+N-1} \right) (N-1)
\end{eqnarray}
When $Z=0$ the bound corresponds to BJB pessimistic bound 

Level 2 pessimistic bound for $Z=0$ is

\begin{eqnarray}
R^{(2)}_{pess} (N) = 1 + \left( \frac{L_b^2 (N-2) }{1 + L_b (N-2)}\right) (N-1)
\end{eqnarray}

A measure of the error magnitude is 

\begin{eqnarray}
\frac{ R^{(i)} (N) - R^{(i)} (N)  }
{ R^{(i)} (N) + R^{(i)} (N) } \times 100\%
\end{eqnarray}

Eager and Sevcik 1986 \cite{EaSe86} is an extension of 
the single class PBH bounding algorithm to QNs with multiple classes of customers. 
The resulting algorithm is applicable to all of the types of product-form QNs 
that are commonly used in computer system and computer-communication network applications. 
{\it Asymptotic Expansions - AE} algorithm proposed by McKenna and Mitra 1994 \cite{McMi84}
is also capable of providing reasonably tight bounds for a practically useful subset of multiple class closed QNs. 
AE algorithm can only treat QNs in which each class has a significant service demand at one or more IS servers. 
Both algorithms offer a hierarchy of bounds with differing accuracy levels and computational cost requirements.

\section*{Appendix II: Kriz's Extension to BJB}\label{sec:kriz}

The discussion follows Kriz 1984 \cite{Kriz84} and preserves its notation. 
The number of jobs is $N$
Single-server or waiting stations are denoted by (${\cal W}$). 
IS or delay stations are denoted ${\cal D}$ and the sum of their service demands by $Z$.
The set of all servers is denoted by ${\cal Q}$ with $M= |{\cal Q}|$.
The loading at server $m$ is $t_m = v_m s_m$,
where $v_m$ is the number of visits and $s_m$ the service time.
The sum of all loading at the ${\cal W}$ servers is $T$ and the sum of all loadings is $r=Z+T$.
The maximum and average loading at ${\cal W}$ servers is $t'$ and $t"$. 

$L_m(N)$ is the queue-length, $R_m(N)$ the response time, 
and $X_m(N)$ the throughput at $m^{th}$ station.
The system throughput is $X(N)=X_m(N)/v_m$ and 
the mean residence time in the system is $R(N) = N/X(N)$.

The throughput ABA and BJB can be expressed succinctly as:

$$X(N) \leq \mbox{min}(N/r, 1/t')$$

$$ \frac{N}{r+(N-1)t'} \leq X(N) \leq \frac{N}{r+(N-1)t"}$$

Given that that load at ${\cal W}$ stations is balanced with loadings given by $t$, 

$$X(1)=1/r, \hspace{3mm} X(n)=  n / [r+(n-1-Z X(n-1)t],  n \geq 2$$

With initialization $\underbar{Y}_0 (N)  = 0$ we have: 

$$\underbar{Y}_i (N)  = N [ r+(N-1- Z Y_{i-1}(N-1))t' ]^{-1}, 1 \leq i \leq N$$

Given $\bar{Y}_0 (N) = \mbox{min} \{ N/r,1/t' \} $

$$\bar{Y}_i (N)= \mbox{min} \left( N [r+(N-1-Z\bar{Y}_{i-1}(N-1)t" ]^{-1} , 1/t' \right)$$

Then $ \bar{Y}_{i-1} (N) \leq
\bar{Y}_{i-1} (N)             \leq
X(N)                          \leq
\bar{Y}_{i} (N)               \leq
\bar{Y}_{i-1} (N)$

\section*{Appendix III: Asymptotic Expansions with Multiple Bottlenecks}\label{sec:George} 

Considered in George et al. 2012 \cite{GeXS12} is a closed QN with $N$ jobs and $M$ stations 
with ${\cal M}$ denoting the set of stations.
${\cal S}$ denotes single-servers, ${\cal L}$ multiservers, and ${\cal I}$ IS stations.
Given the routing matrix $P=[p_{i,j}]$, the relative throughputs are: 
$$\pi = \sum_{j \in {\cal M}} \pi_j p_{j,i}.$$

For any sequence $\{ f(n), n \geq 0 \}$ its z-transform in alternate form is:

$$\tilde{f} (z)  = \sum_{n=0}^\infty f(n)z^{-n} $$

Given three theorems regarding the properties of $\tilde{f}(z)$ 
and an assumption is followed by a theorem. 

Assumption 1. We assume that the relative throughputs $\pi_i, \in {\cal M}$
are chosen such that $\mbox{max} \{\rho_i=1 \}$
stations are labeled by their relative utilization, such that:
$$1=\rho_1 \geq \rho_2 \geq \ldots \geq \rho_M$$

The set of bottleneck stations is defined as ${\cal B} := \{i \in {\cal M}: \rho_i=1\}$.
Clearly $|B| \geq 1$.

The main theorem that characterizes the asymptotic behavior of the 
normalization constant $G(N)$ in exact order.

Theorem 4. For any $M$ station closed QN with $N$ jobs the normalization constant
$G(N)$ satisfies the exact asymptotics:

\begin{eqnarray}\nonumber
G(N) \sim C_B N^{|B|-1}\mbox{   where  }
\end{eqnarray}
\begin{eqnarray}
C_B= \frac{1}{(|B|-1)!}
\prod_{i=1}^{|B|} \frac{\gamma_i^{s^i}}{s_i!} \prod_{j=|B|+1}^M \tilde{f}_j (1)
\end{eqnarray}

As $N$ grows the rate at which $G(N)$ grows is on the order $N^{|B|-1}$.
The actual throughput $\Lambda_i (N)$ and utilization $U_i (m)$ satisfy
the following exact-order asymptotics:

$$\Lambda_i (N) \sim \pi_i (1 - \frac{1}{N})^{|B|-1}, \forall{i} \in {\cal M} $$

$$U_i (N) \sim \rho_i (1 - \frac{1}{N}) ^{|B|-1}, \forall{i} \in {\cal M'}$$

where ${\cal M}$ is the set of all stations and ${\cal M'}$ excludes IS stations (${cal I}$). 

An upper bound to system throughput is given as:

$$\Lambda(N) \leq \mbox{min} \left[ \frac{N}{D+Z}, s_1 \mu_1 \right] := \Lambda^{ABA} (N) $$ 

where $D=\sum_{i \in \cal M'} \pi_i / \pi_1 \mu_i $ and
$Z = \sum_{i \in {\cal I}}  \pi_i / \pi_1 \mu_i$.

As $N$ increases a saturation point is reached where the bottleneck resource reaches 100\% utilization.

Noting that $\Lambda_i (N) = \pi_i / \pi_1 \lambda(N)$

\begin{eqnarray}\label{eq:new}
\mbox{min} \left[ \frac{N}{D+Z}, 
s_1 \mu_1 (1 - \frac{1}{N} )^{|B|-1} \right]
\end{eqnarray}

Numerical examples are given why the AE bound is more accurate than others.

We note that in the case of a network with a single bottleneck, Eq. (\ref{eq:new}) reduces to the previous
ABA bound and the proposed approximation provides the greatest improvements 

\subsection*{Appendix IV: Proportional and Geometric Bounds}\label{sec:app3}

The Geometric Bounds - GBs proposed by Casale et al. 2006/08 \cite{CaMS06,CaMS08}
are fast and accurate noniterative bounds on closed QN metrics. 
Compared to BJB GB achieves higher accuracy at similar computational cost, 
limiting the worst-case bounding error typically within 5-13\% 
while for BJB the error is usually in the range of 15-35\%. 
\footnote{Application of GB as an approximation 
to fork-join processing in closed QNs is beyond the scope of this discussion.}

We restate the steps for MVA \cite{ReLa80} in a closed QN with $M$ queues 
with service demands or loadings $L_i, 1 \leq i \leq M$ with $L=\sum_{i=1}^M L_i $.
The delay at the delay server is $Z$.
The network throughout with $N$ jobs is $X(N)$.
The utilization at the $i^{th}$ queue is $U_i (N)$ and the queue-length $Q_i(N)$.
It follows from Little's result

$$U_i(N) = L_i X(N), 1 \leq i  \leq M$$

According to the arrival theorem \cite{LaRe80} 
the mean residence time at the $i^{th}$ station is:

$$W_i (N) = L_i [ 1 + Q_i (N-1) ], 1 \leq i \leq M $$

The residence time in the network is: 

$$R(N) = \sum_{i=1}^M W_i (N)$$

The network throughput follows from Little's result:

$$X(N) = N / (Z+ R(N)$$

The mean number of jobs at the $i^{th}$ queue is:

$$Q_i = X(N) W_i(N) = U_i (N)  [1+ Q_i (N-1)], 1 \leq i \leq  M$$
\footnote{The paper uses $R_i(N)$ which is undefined, it should be $W_i(N)$.}  

By recursively expanding the above equation:

\begin{eqnarray}\nonumber
Q_i (N) = U_i(N) + U_i(N) U_i (N-1) +  \\
\nonumber
U_i (N) U_i (N-1) U_i (N-2) + \\  
\nonumber
\prod_{i=0}^{N-1} U_i(N-n) 
\end{eqnarray}

The geometric summation yields:

\begin{eqnarray}\label{eq:7}
\sum_{i=1}^N y^i  = \frac{y(1-y^{N})}{1-y}
\end{eqnarray}

The throughput is bounded as follows where $X_{max}$ is the maximum throughput obtained by ABA:

\begin{eqnarray}
N / ( Z+ L N) \leq X(N) \leq                 \\
\nonumber
\mbox{min} (N/(Z+L) , X_{Max})   \mbox{ ABA }
\end{eqnarray}

The throughput for BJBs is:

\begin{eqnarray}
N / (Z+ L + L_{max}  (N-1 -Z X^{-1} ) \leq X(N) \leq        \\
\nonumber
N/ (Z + L + L (N-1) Z X^+ )/M)  \mbox{ BJB } 
\end{eqnarray}

The throughput for {\it Proportional Bounds - PB} Hsieh and Lam 1987 \cite{HsLa87}. 

\begin{eqnarray}
N / \left( 
Z + L  + \frac{ \sum_{i=1}^M  L_i^N ( N - 1 - Z X^- ) } { \sum_{j=1}^M  L_j^{N-1} }  
\right) 
\leq 
X(N)  \\  
\nonumber
\leq
N / 
\left( Z + L  + \frac{ \sum_{i=1}^N L_i^2 ( N-1 - ZX^+) }
{ \sum_{j=1}^M L_j } \right) \mbox{ PB }
\end{eqnarray}

BJB always offers greater accuracy than the ABA.
The bounds hold true for any $X^+$ and $X^-$ such that
$$ X(N-1) \leq X^+ \leq X_{max}, \hspace{5mm} 0 \leq X^-  \leq X(N-1)$$ 

The following formula is exact:

\begin{eqnarray}\label{eq:16}
X(N) = N / [ Z + L+ L_{max} (N-1) - Z X(N-1)) - D(N)]
\end{eqnarray}

where
$$D(N) = \sum_{i=1}^N (L_{max}-L_i) Q_i (N-1).$$ 

{\bf Theorem 4.} $X(n)$ for $Z>0$ is bounded by

\begin{eqnarray}
2N / 
\left( b(N) + \sqrt{b^2 (N)  - 4 Z L_{max} (N-1)} \right) 
\leq X(N) \leq                                                        \\
\nonumber
2N / \left( b(N) + \sqrt{b^2 (N)  - 4 Z L_{max} N } \right)  
\end{eqnarray}

\begin{eqnarray}
b(N) = Z + L + L_{max} (N-1) - \sum_{i:L_i <  L_{max}} \\
\nonumber
(L_{max} - L_i) Q_i (N-1).
\end{eqnarray}

$Q_i$ is bounded by $Q_i^-$ and $Q_i^+$ given 
by Eq. (\ref{eq:1} and Eq. (\ref{eq:2}), respectively.

{\bf Theorem 1.} The queue length of station is bounded from below by 

\begin{eqnarray}\label{eq:1}
Q_i^- (N) =
\begin{cases}
\frac{ y_i (N) - {y_i (N)}^{N+1} }{ 1- y_i (N) }  \mbox{ if } L_i < L_{max},  \\        
\nonumber
\frac{1}{m_{max}} 
\left( N- Z X^+ - \sum_{k:L_k < M_{max} } Q^+_k (N)  \right)   \\
\nonumber
\mbox{ if } L_i = L_{max}
\end{cases}
\end{eqnarray}

For any $X^+$ such that $X(N) < X^+$ and where
$$ y_i(N) = L_i N / (Z+ L+ L_{max} N) $$
is the ratio of the underlying sum in Eq. (\ref{eq:7}),
$M_{max}$ is the number of queues with service demand $L_{max}$,
and $Q^+_k$ is the upper bound in Theorem 2 for $L_i < L_{max}$

{\bf Theorem 2.} The queue-length $Q_i (N)$ is bounded above as follows:

\begin{eqnarray}\label{eq:2}
Q^+_i (N) =
\begin{cases}
\frac { Y_i (N) [1  - {Y_i (N) }^N ] } {1- Y_i (N)} \mbox{  if  }L_i < L_{max}   \\
\nonumber
\frac{1}{ M_{max} } \left( N - Z  X^- - \sum_{k:L_k < L_{max}} Q^+_k (N) \right) \\
\nonumber
\mbox{  if  }L_i = L_{max}.   
\end{cases}
\end{eqnarray}

For any $X^+$ and $X^-$ such that $X(N) \leq X^+  \leq X_{max} $ and $ 0  \leq X^- \leq X(N) $
and where $Y_i (N) = L_i^+$ is the ratio $y$ of the underlying geometric sum.
$$ \frac{N-1}{N} X(N) \leq X(N-1) \leq X(N)$$
$X^-$ and $X^+$ are used to represent lower and upper bounds.

The {it Geometric Square Bound - GSB} is obtained by replacing queue=lengths by $Q_i^- (N-1)$.

Two classes of performance bounds one for single chain
and the other for multi-chain QNs are proposed in \cite{HsLa87}.
{\it Proportional Bounds - PBs} assume that mean queue-lengths are proportional to server loads.
PBs are more accurate than BJBs in that individual server loads
are retained as parameters in bounds formula.
Several tables are given in \cite{CaMS08} comparing the accuracy of various methods.
GSB is shown to be the most accurate even when one or two iterations are allowed for BJB and PB. 

\section*{Appendix V: Proportional Approximation Method - PAM}\label{sec:app4}

{\it Proportional Approximation Method - PAM} is that latest bounding method 
for closed multichain QNs by Hsieh and Lam 1988/89 \cite{HsLa88,HsLa89}.

Approximate MVA algorithms for separable queueing networks 
are based upon an iterative solution of a set of modified MVA formulas. 
Let $M$ denotes the number of queues and $K$ the number of chains or job classes.
Each iteration has a computational cost of O($MK^2$) or less, 
many iterations are typically needed to attain convergence. 

Presented are faster approximate noniterative solution algorithms,
which are suitable for the analysis and 
design of communication networks requiring thousands of closed chains.

Three PAM algorithms of increasing accuracy are presented. 
Two of them have time and space requirements of O($MK$). 
The third algorithm has a time requirement of O($MK^2$) and a space requirement of O($MK$).

Three PAMs developed in this study consider QNs with fixed-rate and delay servers.

Let $\tau_{mk}$ denote the loading of chain $k$ at queue $m$.

{\bf Algorithm PAM\_BASIC}

{\bf Step 1.} Calculate proportional approximations of mean queue lengths.

$$\gamma_{mk} = \tau_{mk} / \sum_{i=1}^M \tau_{ik}$$

$$\mbox{for }m=1, \ldots, M, \mbox{  and  }k=1, \ldots, K$$

$${q'}_{mh} ( \underbar{N}) = \gamma_{nh} N_h  $$

$${q'}_{mh} (\underbar{N} - \underbar{1}_K) = 
\begin{cases}
{q'}_{mh} (\underbar{N})\mbox{ if  } h\neq k. \\
{q'}_{mh} (\underbar{N})-\gamma_{mh} \mbox{ if  } h = k. \\
\end{cases}
$$

$$\mbox{for }m=1, \ldots, M, \mbox{  and  }k=1, \ldots, K$$

{\bf Step 2:} Calculate approximate mean delay of chain $k$
at server $m$ and approximate throughput of chain $k$

$$
D_{mk} (\underbar{N}) =
\begin{cases}
\tau_{mk} [ 1 + \sum_{k=1}^K {q'}_{mh} (\underbar{N})  - \underbar{1}_k) ] \mbox{ if fixed rate}  \\
\tau_{mk} \mbox{ if delay server}
\end{cases}
$$

$$T_k(\underbar{N}) = \frac{N_k}{\sum_{m=1}^M  D_{mk} (\underbar{N})} \mbox{ for }k=1, \ldots K$$ 

The accuracy improvement of PAM\_IMPROVED over PAM\_BASIC is obtained 
by a simple scaling operation to ensure that server utilizations do not exceed one. 

Algorithm PAM\_Improved calculates server utilizations and 
finds the the server with maximum utilization visited by chain $k$.

{\bf Step 3.}  

$$U_m (\underbar{N}) = \sum_{k=1}^K T_k (\underbar{N}),\hspace{3mm}m=1,\ldots,M$$

{\bf Step 4.} Find the largest utilization $S_k$ visited by chain $k$.

$$ S_k = \mbox{max} \{U_m (\underbar{N}) \}\hspace{3mm}k=1, \ldots, K$$

{\bf Step 5.} Scale down the throughputs of individual chains in necessary 

for $k=1, \ldots, K $

$$\mbox{ If }S_k  > 1\mbox{ then } T_k(\underbar{N}) = T_k (\underbar{N})/S_k, $$ 

{\bf Step 6.} Calculate total throughput and utilizations 

$$T(\underbar{N}) = \sum_{k=1}^K T_k (\underbar{N})$$
 
$$ U_m (\underbar{N}) = \sum_{k=1}^ K \tau_{ml} T_k (\underbar{N}), \hspace{3mm}m=1,\ldots,M$$

{\bf Algorithm PAM\_II}

The accuracy of the PAM algorithm is improved
by executing the last two steps of the MVA recursion
instead of just the last step using the proportional approximation 
to get initial mean queue length estimates. 
The reader is referred to \cite{HsLa88,HsLa89}
Relative errors in chain throughputs for 500 networks
calculated by PAM\_Improved and PAM\_TWO are 2.3\% vs 0.8\% 
for average error and 40.3 vs 30.8\% for the maximum.

The additional accuracy of PAM\_TWO is obtained by executing 
the final two steps of the MVA recursion instead of just the last step.
The computational time requirements are O(MK) for PAM\_BASIC and PAM\_IMPROVED, 
and O($M^2$) for PAM\_TWO. 
All three PAM algorithms have space requirements of O(MK).

\subsection*{Acknowledgement}
The discussion of UJA is based on joint publications 
with Dr Behzad (Brad) Nadji at the EE-Systems Dept. at USC.


\end{document}